\documentclass[12pt,a4,pdftex]{article}

\usepackage[latin1]{inputenc}
\usepackage{amssymb,amsmath,amsthm,float,amsfonts,makeidx,lscape,subfigure}
\usepackage{rotating,multirow,multicol,colortbl,supertabular}
\usepackage[hyphens]{url}
\usepackage{color}
\usepackage[colorlinks=true,linkcolor=blue,citecolor=blue,filecolor=blue]{hyperref}
\usepackage{lscape}

\usepackage{url}
\DeclareUrlCommand\url{\color{blue}}

\usepackage[driver=pdftex]{geometry}

\newif\ifpdf
\ifx\pdfoutput\undefined
  \pdffalse
\else
  \pdfoutput=1
  \pdftrue
\fi

\usepackage{harvard}
\makeatletter
\renewcommand{\@seccntformat}[1]{{\csname the#1\endcsname}.\hspace{0.5em}}

\long\def\@makefigcaption#1#2{%
  \vskip\abovecaptionskip
  \sbox\@tempboxa{\textbf{#1.} #2}%
  \ifdim \wd\@tempboxa >\hsize
  \textbf{#1.} #2\par
  \else
    \global \@minipagefalse
    \hb@xt@\hsize{\hfil\box\@tempboxa\hfil}%
  \fi
  \vskip\belowcaptionskip}

\renewcommand{\figure}{\let\@makecaption\@makefigcaption\@float{figure}}

\long\def\@maketblcaption#1#2{%
  \vskip\abovecaptionskip
  \small{\bf#1. }\normalsize#2\vspace*{1.5mm}
  \vskip\belowcaptionskip}

\renewcommand{\table}{\let\@makecaption\@maketblcaption\@float{table}}

\makeatother

\usepackage{setspace}
\makeatletter
\def\needspc#1{\vskip\z@ plus #1 \penalty-60 \vskip \z@ plus-#1\relax}
\def\@listi{\topsep 0.0ex \parsep 0pt \leftmargin 1cm}
\let\@listI\@listi
\makeatother

\newcommand{\be}{\begin{equation}}
\newcommand{\ee}{\end{equation}}
\newtheorem{theorem}{Theorem}

\newcommand{\nothing}[1]{}

\begin{document}

\thispagestyle{empty}

\begin{center}

\LARGE Spurious Default Probability Projections in \\ Credit Risk Stress Testing Models

\vspace*{1.5cm}

\large

Bernd Engelmann\ \footnote{\ Ho Chi Minh City Open University, 35-37 Ho Hao Hon Street, District 1, Ho Chi Minh City, e-mail: \texttt{\href{mailto:bernd.engelmann@ou.edu.vn}{\color{blue}bernd.engelmann@ou.edu.vn}}}\\

\vspace*{0.75cm}

This version: January 16, 2024

\end{center}

\normalsize

\vspace*{0.75cm}

\begin{abstract}
\begin{spacing}{1.0}
Credit risk stress testing has become an important risk management device which is used both by banks internally and by regulators. Stress testing is complex because it essentially means projecting a bank's full balance sheet conditional on a macroeconomic scenario over multiple years. Part of the complexity stems from using a wide range of model parameters for, e.g., rating transition, write-off rules, prepayment, or origination of new loans. A typical parameterization of a credit risk stress test model specifies parameters linked to an average economic, the through-the-cycle, state. These parameters are transformed to a stressed state by utilizing a macroeconomic model. It will be shown that the model parameterization implies a unique through-the-cycle portfolio which is unrelated to a bank's current portfolio. Independent of the stress imposed to the model, the current portfolio will have a tendency to propagate towards the through-the-cycle portfolio. This could create unwanted spurious effects on projected portfolio default rates especially when a stress test model's parameterization is inconsistent with a bank's current portfolio.
\end{spacing}
\end{abstract}

\begin{spacing}{1.0}
\parindent0pt {\sl \bf JEL Classification:} G21\\
\parindent0pt {\sl \bf Keywords:} Credit Risk, Stress Testing, Default Probability, Transition Matrix, Spurious Projections
\end{spacing}

\clearpage

\pagestyle{plain}

\addtocounter{page}{-1}

\section{Introduction}\label{sec:intro}

Stress testing has become a core part of banks' risk management in the past two decades. Regulators require banks to perform stress tests regularly \cite{bas:18} where a bank has to define stress scenarios relevant to its business internally. In addition, banks have to perform stress tests on request of regulators where scenarios and instructions are provided by regulating authorities. In the US, the Federal Reserve Bank (FED) performs stress test for the largest US banks regularly\footnote{\ See {\href{https://www.federalreserve.gov/supervisionreg/stress-tests-capital-planning.htm}{\color{blue}https://www.federalreserve.gov/supervisionreg/stress-tests-capital-planning.htm}}} while in Europe, this is done by the European Banking Authority (EBA).\footnote{\ See {\href{https://www.eba.europa.eu/risk-analysis-and-data/eu-wide-stress-testing}{\color{blue}https://www.eba.europa.eu/risk-analysis-and-data/eu-wide-stress-testing}}} For most of these stress tests, the projection time horizon was three years. In the last two years, a new form of regulatory stress tests emerged which aims at stress testing bank portfolios under climate scenarios. A key difference to previous stress tests is the much longer time horizon which could be up to 30 years. Examples are the 2021 stress test of the Bank of England\footnote{\ See {\href{https://www.bankofengland.co.uk/stress-testing/2021/key-elements-2021-biennial-exploratory-scenario-financial-risks-climate-change}{\color{blue}https://www.bankofengland.co.uk/stress-testing/2021/key-elements-2021-biennial-exploratory-sc enario-financial-risks-climate-change}}}, the 2022 EBA climate stress test \cite{ecb:22}, the 2023 stress test of the Reserve Bank of New Zealand (RBNZ)\footnote{\ See {\href{https://www.rbnz.govt.nz/financial-stability/stress-testing-regulated-entities/climate-stress-test}{\color{blue}https://www.rbnz.govt.nz/financial-stability/stress-testing-regulated-entities/climate-stress-test}}}, or the 2023 climate scenario analysis of the FED.\footnote{\ See {\href{https://www.federalreserve.gov/publications/climate-scenario-analysis-exercise-instructions.htm}{\color{blue}https://www.federalreserve.gov/publications/climate-scenario-analysis-exercise-instructions.htm}}} The focus of regulatory stress tests is on credit risk in banks' loan portfolios which is the dominant risk category for most banks. For this reason, this article will concentrate on credit risk stress testing and leave other risks aside.

Although there is no standard model or modeling framework for a credit risk stress test implementation, most banks follow a common modeling philosophy. The core parameters controlling the credit risk of a loan portfolio are default probabilities (PD), loss given default (LGD) and exposure at default (EAD). Usually migration between rating categories is allowed and modeled by transition matrices. These core risk parameters are provided for an average state of the economy. They are also known as through-the-cycle (TTC) risk parameters. Stress scenarios are typically provided by macroeconomic scenarios and a macroeconomic model is translating the scenario into an abstract factor representing the state of the economy. This factor is then used to transform TTC risk parameters into point-in-time (PIT) or stressed risk parameters which are used to evaluate the impact of the stress scenario on bank capital, loan loss provisions, and interest income.

When the stress horizon is long as in the aforementioned stress tests focusing on climate risk scenarios, it is not realistic to work with a static portfolio but assumptions on write-off, run-off, prepayment, and new origination have to be made to achieve a realistic portfolio propagation. It turns out that these parameters together with the TTC risk parameters define a TTC portfolio that is independent of the current loan portfolio of a bank. Furthermore, the current portfolio has a tendency to propagate towards this TTC portfolio. When the parameters used in the stress test model are inconsistent with the current portfolio, spurious projections of the average PD of a loan portfolio might occur that are hidden in the stress test parameterization and distort the stress test outcome. A typical situation where this could occur is when a bank is using external data due to the lack of internal data. An example could be a bank's large corporate portfolio where the bank might be unable to estimate an internal rating transition matrix and is instead relying on a matrix published by a rating agency.

This article is organized as follows. In Section \ref{sec:model}, a stress testing model is outlined that is representative of what is used by many banks and serves as a reference for this article. In Section \ref{sec:ttc}, the TTC portfolio is characterized, the conditions under which a unique TTC portfolio exists are derived, and an algorithm for computing the TTC portfolio is provided. In Section \ref{sec:examples}, examples are calculated for illustration and a graphical method for easily detecting spurious PD projections is given. The final section concludes.

\section{Credit Risk Stress Testing Model}\label{sec:model}

Credit portfolio modeling started in the 1990s when models to measure credit portfolio risk by risk measures like value-at-risk or expected shortfall have been developed. The most popular of these models is CreditMetrics \cite{gup:fin:bha:97}. The modeling framework developed in CreditMetrics is the core building block of both the Basel II capital functions \cite{gordy2003risk} and many stress testing models applied today.

The starting point of this modeling framework is a one-period one-factor model for credit risk:
\begin{equation}
	r = \sqrt{\rho} Z + \sqrt{1 - \rho}\epsilon,
\end{equation}
where $r$ is the log-return of a borrower's assets, $Z$ a random systemic factor common to all borrowers, $\epsilon$ a borrower-specific random factor, and $\rho$ the correlation between the asset log-returns of two borrowers. Both random variables $Z$ and $\epsilon$ are assumed to be standard normally distributed and independent. Furthermore, borrower-specific random variables $\epsilon$ are assumed to be independent between borrowers. A borrower defaults when $r$ falls below a threshold $\theta$. By construction $r$ is standard normally distributed which allows a characterization of $\theta$ in terms of the unconditional borrower default probability $p$:
\begin{equation}
	p = P\left(r < \theta\right) = \Phi\left(\theta\right) \Rightarrow \theta = \Phi^{-1}\left(p\right),
\end{equation}
where $\Phi$ is the cumulative distribution function of the standard normal distribution.

The systemic factor $Z$ could be viewed as an abstract representation of the state of the economy. A positive realization $z$ of $Z$ represents a boom and reduces the likelihood of a borrower's asset return falling below $\theta$ conditional on $z$. A negative realization of $z$, on the contrary, models a recession and increases default probabilities. The borrower default probability conditional on $z$ is denoted with $p(z)$ and computed as
\begin{equation}\label{eq:pit:ttc}
	p(z) = P\left(r < \theta\right) = P\left(\epsilon < \frac{\theta - \sqrt{\rho} z}{\sqrt{1 - \rho}}\right) = \Phi\left(\frac{\Phi^{-1}\left(p\right) - \sqrt{\rho} z}{\sqrt{1 - \rho}}\right)
\end{equation}
The relation between $p(z)$ and $p$ in (\ref{eq:pit:ttc}) is essentially a transformation between a PIT default probability conditional on the state of the economy $z$ and an unconditional TTC default probability. This relation has been exploited in the construction of PIT-TTC-PD frameworks for regulatory and internal risk management purposes \cite{aguaisdesigning,car:pet:12}. In a stress test, typically the TTC parameter is an input, the state of the economy $z$ is derived from macroeconomic models, and (\ref{eq:pit:ttc}) is used to compute risk parameters that are needed to evaluate the scenario impact on a bank's portfolio.

In \citeasnoun{gup:fin:bha:97}, an extension of (\ref{eq:pit:ttc}) to transition matrices is developed. The starting point is a TTC transition matrix ${\mathbf T}$ which is defined as
\begin{equation}
	{\mathbf T} = \left(\begin{array}{ccccc} p_{1,1} & p_{1,2} & \cdots & p_{1,n-1} & p_{1,n} \\
							     p_{2,1} & p_{2,2} & \cdots & p_{2,n-1} & p_{2,n} \\
							     \vdots   & \vdots   & \ddots & \vdots & \vdots   \\
							     p_{n-1,1} & p_{n-1,2} & \cdots & p_{n-1,n-1} & p_{n-1,n} \\
								0      &    0   &    \cdots & 0 & 1  \end{array}\right)\ ,
\end{equation}
where $n$ is the number of rating grades and $p_{i,j}$ the probability that a borrower in rating grade $i$ moves within one year into rating grade $j$. The $n^{th}$ grade is the default grade. All rows in ${\mathbf T}$ sum to one, i.e., $\sum_{j=1}^np_{i,j} = 1$. The default grade is assumed to be absorbing which means cures are not possible under the migrations implied by ${\mathbf T}$. To compute a transition matrix conditional on the state of the economy $z$, for each rating grade $i$, $n-1$ thresholds $\theta_{i,j}$ are introduced. A borrower in rating $i$ migrates to rating $j$ if the log-return of his assets $r$ fulfills $\theta_{i,j} \le r < \theta_{i,j-1}$. To compute ${\mathbf T}(z) = \left(p_{i,j}(z)\right)$, a generalization of (\ref{eq:pit:ttc}) is needed:
\begin{equation}\label{eq:stress:tm}
	p_{i,j}(z) = \left\{\begin{array}{ll} \Phi\left(\frac{\Phi^{-1}\left(p_{i,n}\right) - \sqrt{\rho} z}{\sqrt{1 - \rho}}\right)  & ,\ \mbox{if}\ j=n \\
\Phi\left(\frac{\Phi^{-1}\left(\sum_{l=j}^np_{i,l}\right) - \sqrt{\rho} z}{\sqrt{1 - \rho}}\right) - \sum_{l=j+1}^np_{i,j}(z)  & ,\ \mbox{if}\ j=2,\ldots,n-1 \\ 1 - \sum_{l=2}^np_{i,j}(z)  & ,\ \mbox{if}\ j=1 \end{array}\right.
\end{equation}
To apply (\ref{eq:stress:tm}) in a stress testing context, a stressed value $z_{stress}$ of the state of the economy has to be provided to transform an average into a stressed transition matrix. The stressed transition matrix is then applied on a bank's portfolio to determine the impact of a stress scenario. This approach was applied, e.g., in \citeasnoun{bangia2002ratings}, \citeasnoun{de2013stress}, \citeasnoun{miu2009stress}, or \citeasnoun{witzany2020stressing}.

The stressed state of the economy $z_{stress}$ is usually derived from a macroeconomic model. The basis of a macroeconomic model is a credit index $C$ capturing the economic cycle. This could be the time series of a default rate of a bank's portfolio or some externally published default or bankruptcy rate. The macroeconomic model is building a link between the time series of the credit index $C_t$ and macroeconomic variables $X_{k,t}$ like the unemployment rate, GDP growth or other variables that are suitable for explaining the behavior of $C$. Macroeconomic models are widely applied in risk management also for other purposes than stress testing, e.g., for default probability projections when calculating ECL under IFRS 9 \cite{pesaran2006macroeconomic,schechtman2012macro,skoglund2016application}.

To derive a stressed state of the economy $z_{stress}$ from a macroeconomic scenario, the starting point is a macroeconomic model, e.g.,
\begin{equation}\label{eq:macro}
	\Phi^{-1}\left(C_t\right) = \beta_0 + \beta_1X_{1,t-l} + \ldots + \beta_kX_{k,t-l},
\end{equation}
where $l$ is the time lag used in the model estimation. The historically observed credit index $C_t$ could be viewed as a realization of a PIT default probability $p(z)$. Utilizing (\ref{eq:pit:ttc}) under this interpretation of $C_t$ leads to a connection between $C_t$ and $z_t$
\begin{equation}\label{eq:soc}
	\Phi^{-1}\left(C_t\right) = \frac{\Phi^{-1}\left(p\right) - \sqrt{\rho} z_t}{\sqrt{1 - \rho}}
\end{equation}
The parameters $p$ and $\rho$ can be estimated from the times series $C_t$. This could be done either by the method of moments used in \citeasnoun{car:pet:12} or a more advanced method described in \citeasnoun{duellmann2010estimating}. Once the parameters $p$ and $\rho$ are estimated, a link between macroeconomic variables and the state of the economy $z_t$ could be established easily by combining (\ref{eq:macro}) and (\ref{eq:soc})
\begin{equation}\label{eq:stress}
	z_t = \frac{\sqrt{1-\rho}\left(\beta_0 + \beta_1X_{1,t-l} + \ldots + \beta_kX_{k,t-l}\right)-p}{\sqrt{\rho}}.
\end{equation}
A detailed worked-out example using this approach for residential mortgages with data from a CCAR stress test of the US FED could be found in \citeasnoun{be2021basel}.

In this stress testing framework, a macroeconomic stress scenario is provided, transformed by (\ref{eq:stress}) into a stressed state of the economy $z_{stress}$ which is then used to transform a TTC transition matrix into a stressed transition matrix by means of (\ref{eq:stress:tm}). Ideally, the economic stress is entirely captured in $z$ while the role of the transition matrix is distributing loan balance appropriately among rating grades. However, as will be shown in the next two sections, the transition matrix may have "an own life" which distorts the stress test and could lead to spurious outcomes.

\section{Through-the-Cycle Portfolio}\label{sec:ttc}

So far, a core general framework for credit risk stress testing was outlined. Although concrete implementations in different banks might differ in multiple aspects, the general framework is still a reasonable starting point for analyzing stress testing models. To study the dynamics implied by a stress test model's parameterization, the following model structure is assumed:
\begin{enumerate}
\item A portfolio has $n$ rating grades and its initial distribution is $W = \left(w_1,w_2,\ldots,w_n\right)$. The distribution is given as percentages and it holds $\sum_{i=1}^nw_i = 1$
\item In addition to the initial portfolio, a TTC transition matrix ${\mathbf T}$ and an origination vector $O = \left(o_1,o_2,\ldots,o_n\right)$ is provided as input
\item A time series for the state of the economy $z_t$, $t=1,\ldots,m$ is estimated from a macroeconomic model like in (\ref{eq:stress})
\item In each period t, ${\mathbf T}(z_t)$ is computed and the portfolio $W_{t-1}$ is propagated using ${\mathbf T}(z_t)$ to arrive at $W_t$
\item The defaulted part of the portfolio after migration $w_{t,n}$ is immediately written off and replaced by new origination of exactly the defaulted amount to keep total loan balance constant over time. The new origination is controlled by $O$ which gives the percentage of originated loans per rating grade. The condition $\sum_{i=1}^no_i = 1$ holds
\end{enumerate}
First, a full propagation step of the above algorithm will be represented in matrix notation. Assume $t-1$ steps are completed resulting in a portfolio distribution $W_{t-1}$. Then $W_t$ is computed as
\begin{equation}\label{eq:propagation}
	W'_t = W'_{t-1} {\mathbf T}(z_t) {\mathbf I}_w + \left(W'_{t-1} {\mathbf T}(z_t) V_w\right) O',
\end{equation}
where ${\mathbf I}_w$ is a matrix that creates the immediate write off of defaulted exposures
\begin{equation*}
	{\mathbf I}_w = \left(\begin{array}{ccccc} 1 & 0 & \cdots & 0 & 0 \\
							     0 & 1 & \cdots & 0 & 0 \\
							     \vdots   & \vdots   & \ddots & \vdots & \vdots   \\
							     0 & 0 & \cdots & 1 & 0 \\
								0      &    0   &    \cdots & 0 & 0  \end{array}\right)
\end{equation*}
and $V_w = \left(0, 0, \ldots, 0, 1\right)$ is extracting the defaulted balance after migration and distributes it among the $n$ rating grades using $O$. The symbol $'$ indicates a transposed vector or matrix.\\

\begin{theorem}\label{th:ttc}
	If the following two conditions
	\begin{enumerate}
		\item The sub-matrix of transitions between performing rating grades ${\mathbf T}_p = (p_{i,j}),\ i,j = 1,\ldots,n-1$ is primitive, i.e., there exists a positive integer $k_p$ such that all entries of ${\mathbf T}_p^{k}$ are strictly positive $\forall k \ge k_p$
		\item There is no origination into the default class, i.e., $o_n = 0$
	\end{enumerate}
	are fulfilled, then there exists a unique portfolio $W_{ttc}$ with
	\begin{equation}\label{eq:ttc}
		W'_{ttc} = W'_{ttc} {\mathbf T} {\mathbf I}_w + \left(W'_{ttc} {\mathbf T} V_w\right) O'.
	\end{equation}
	The TTC portfolio $W_{ttc}$ can be calculated by the algorithm
	\begin{align*}
		W_0 & = W_{init} \\
		W'_t & = W'_{t-1} {\mathbf T} {\mathbf I}_w + \left(W'_{t-1} {\mathbf T} V_w\right) O',\ t = 1,2,\ldots
	\end{align*}
	for an arbitrary initial portfolio $W_{init}$.
\end{theorem}
\vspace*{2mm}
From a practical perspective, the two conditions of Theorem \ref{th:ttc} are not restrictive. Origination of new loans to defaulted borrowers is aside from relatively rare investments into distressed loans not part of a bank's business. Also Condition 1 should be fulfilled for basically all rating systems. Economically, it states that it must be possible to migrate from every performing rating grade $i$ to any performing rating grade $j$. This migration does not necessarily have to happen within one period, but over multiple periods every rating migration between performing grades must be possible which is reflected in the positiveness of the matrix ${\mathbf T}_p^k$. The proof of Theorem \ref{th:ttc} is provided in the appendix.

Note, if the two conditions are not fulfilled, simple counterexamples, where no TTC portfolio exists, can be constructed. Consider a rating system with $n = 3$ grades, TTC transition matrix
\begin{equation*}
	{\mathbf T} = \left(\begin{array}{ccc} 0 & 1 & 0 \\ 1 & 0 & 0 \\ 0 & 0 & 1 \\ \end{array}\right),
\end{equation*}
current portfolio $W = (1,0,0)$, and arbitrary origination vector $O$. In this case, no defaults occur and, when applying the iterative algorithm, the portfolio switches each year between rating grades 1 and 2 without converging to a TTC portfolio $W_{TTC}$. The reason for the failure to converge is that the sub-matrix between performing grades
\begin{equation*}
	{\mathbf T_p} = \left(\begin{array}{cc} 0 & 1 \\ 1 & 0  \end{array}\right)
\end{equation*}
is not primitive as repeated multiplications of ${\mathbf T_p}$ with itself never lead to matrix with strictly positive entries.

\section{Numerical Examples}\label{sec:examples}

Theorem \ref{th:ttc} states if two mild conditions are fulfilled, there exists a unique portfolio $W_{ttc}$ which depends only on a stress test's model parameters not on the current bank portfolio. Some numerical examples will illustrate the problems that might occur, especially in cases where the transition matrix has not been estimated on the stressed portfolio. This situation could happen regularly in practice. For instance, a bank might use transition matrices provided by rating agencies due to the lack of own data. Another common situation for banks operating in multiple jurisdictions, is to use a transition matrix estimated for a jurisdiction where that bank has a rich data set in a jurisdiction which is considered a similar but where data is scarce. An example could be an Australian bank using model parameters estimated on Australian data for New Zealand portfolios.

To provide some illustrative examples, a transition matrix ${\mathbf T}$ from \citeasnoun{trueck2009rating}, page 3, is used
\begin{equation}
	{\mathbf T} = \left(\begin{array}{cccccccc}
	0.9276 & 0.0662 & 0.0050 & 0.0009 & 0.0003 & 0.0000 & 0.0000 & 0.0000 \\
	0.0064 & 0.9152 & 0.0700 & 0.0062 & 0.0008 & 0.0011 & 0.0002 & 0.0001 \\
	0.0007 & 0.0221 & 0.9137 & 0.0546 & 0.0058 & 0.0024 & 0.0003 & 0.0005 \\
	0.0005 & 0.0029 & 0.0550 & 0.8753 & 0.0506 & 0.0108 & 0.0021 & 0.0029 \\
	0.0002 & 0.0011 & 0.0052 & 0.0712 & 0.8229 & 0.0741 & 0.0111 & 0.0141 \\
	0.0000 & 0.0010 & 0.0035 & 0.0047 & 0.0588 & 0.8323 & 0.0385 & 0.0612 \\
	0.0012 & 0.0000 & 0.0029 & 0.0053 & 0.0157 & 0.1121 & 0.6238 & 0.2389 \\
	0.0000 & 0.0000 & 0.0000 & 0.0000 & 0.0000 & 0.0000 & 0.0000 & 1.0000
	\end{array}\right)
\end{equation}
together with the origination vector $O = (0.00$, $0.20$, $0.30$, $0.30$, $0.20$, $0.00$, $0.00$, $0.00)$. An application of the iterative algorithm of Theorem \ref{th:ttc} leads to the TTC portfolio $W_{ttc} = (0.0183$, $0.1423$, $0.3379$, $0.2633$, $0.1321$, $0.0911$, $0.0150$, $0)$. The corresponding portfolio TTC PD is 1.198\%.

\begin{figure}[!ht]
	\centering
	\includegraphics[scale=0.60]{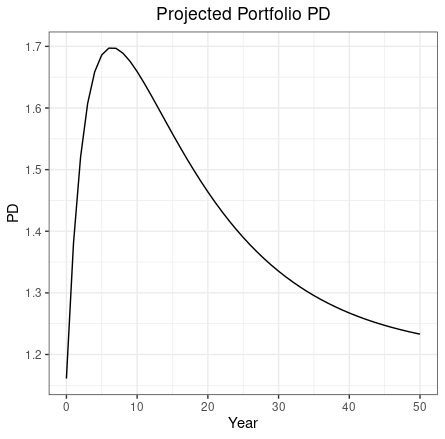}\\
	\caption{Average PD of the projected portfolio $W_{init}$ over 50 years using ${\mathbf T}$ without stress}\label{fig:example1}
\end{figure}

Suppose the current bank portfolio is $W_{init} = (0.00$, $0.00$, $0.20$, $0.40$, $0.30$, $0.10$, $0.00$, $0.00)$. The portfolio PD is 1.161\% which is slightly lower than the PD of the portfolio $W_{ttc}$. If the portfolio $W_{init}$ is propagated over multiple years using ${\mathbf T}$, one would expect from Theorem \ref{th:ttc} that the portfolio $W_{init}$ converges to $W_{ttc}$. The convergence is confirmed in Figure \ref{fig:example1} which displays the average PD of the propagated portfolio over time. However, the convergence is not monotonic. The portfolio $W_{init}$ contains zeros in the best two rating grades which is inconsistent with the rating migration matrix ${\mathbf T}$. This matrix implies that a seasoned portfolio should contain some balance in the highest two rating grades. Although there was no stress applied in the creation of Figure \ref{fig:example1}, the outcome driven by the inconsistency of $W_{init}$ and ${\mathbf T}$ signals a severe recession. Note, that this spurious recession is visible in the first years of the stress test. This means that even the results of three-year projections like in the regular EBA stress tests might be impacted by this effect.\footnote{\ Mathematically, in the language of the proof provided in the appendix, the decomposition of $W_{init}$ into the eigenvectors of the matrix ${\mathbf M}_p$ is dominated by eigenvectors other than $v_1$. These eigenvectors create the spike in the portfolio PD and it takes some time until their impact is dampened by repeated migrations.}

\begin{figure}[!ht]
	\centering
	\includegraphics[scale=0.60]{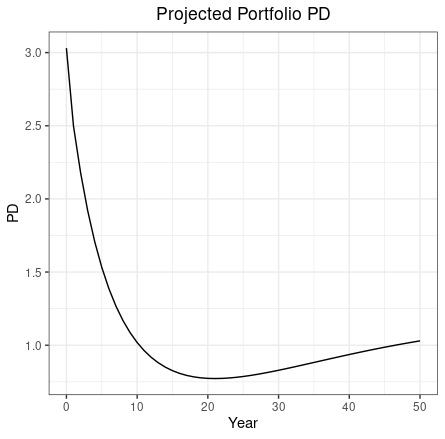}\\
	\caption{Average PD of the projected portfolio $\overline{W}_{init}$ over 50 years using ${\mathbf T}$ without stress}\label{fig:example2}
\end{figure}

The next example is using an even more extreme deviation from $W_{ttc}$, the initial portfolio $\overline{W}_{init} = (0.70$, $0.00$, $0.00$, $0.00$, $0.00$, $0.25$, $0.05$, $0.00)$ which has an average portfolio PD of 2.725\%. The propagation of this portfolio using ${\mathbf T}$ is displayed in Figure \ref{fig:example2}. Again the convergence is not monotonic. Before converging to $W_{ttc}$ the portfolio reaches a minimum average PD of 0.722\% indicating a benign economy although no positive $z$ was applied in the portfolio projection.

\begin{figure}[!ht]
	\centering
	\includegraphics[scale=0.60]{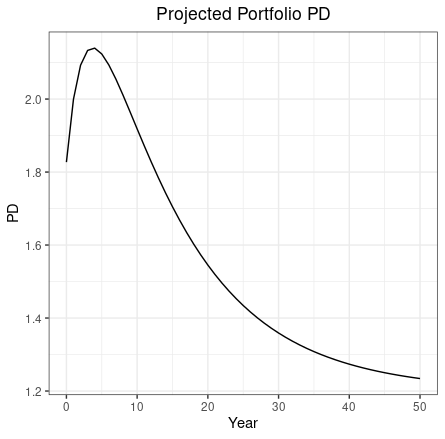}\\
	\caption{Average PD of the projected portfolio $\widetilde{W}_{init}$ over 50 years using ${\mathbf T}$ without stress}\label{fig:example3}
\end{figure}

In Figure \ref{fig:example2}, the portfolio was in recession and PDs moved downwards quickly. This is not necessarily always the case. For the initial portfolio $\widetilde{W} = (0.01$, $0.02$, $0.10$, $0.30$, $0.41$, $0.15$, $0.01$, $0.00)$ with average portfolio PD 1.83\% the projection is shown in Figure \ref{fig:example3}. The portfolio PD moves up to a maximum of 2.14\% before the portfolio starts converging towards $W_{ttc}$. This example also illustrates that zeros in the initial portfolio are not the root cause of the problem but whether the deviation of the initial portfolio from $W_{ttc}$ could be explained by migrations that contribute to the average migration matrix ${\mathbf T}$.

\begin{figure}[!ht]
	\centering
	\includegraphics[scale=0.60]{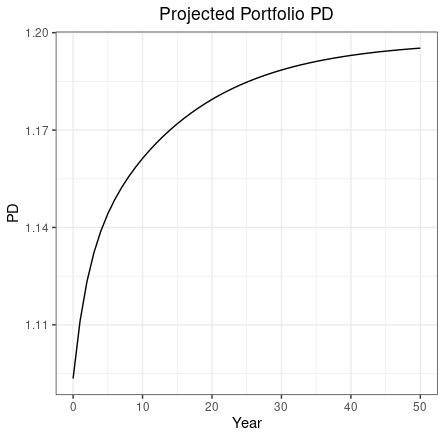}\\
	\caption{Average PD of the projected portfolio $\widehat{W}_{init}$ over 50 years using ${\mathbf T}$ without stress}\label{fig:example4}
\end{figure}

In the final example, the initial portfolio is $\widehat{W}_{init} = (0.0199$, $0.1516$, $0.3472$, $0.2568$, $0.1263$, $0.0857$, $0.0125$, $0.0000)$ with average portfolio PD 1.093\%. The portfolio was constructed by stressing the matrix ${\mathbf T}$ with $z =1$ and performing one propagation step using $W_{ttc}$ as initial portfolio. In this case the initial portfolio is consistent with the credit stress test parameterization and the projection displayed in Figure \ref{fig:example4} is well-behaved.

Although some of the initial portfolios used in this section were rather extreme, the examples illustrate that basically anything could happen when a migration matrix is used that is inconsistent with a bank's portfolio. Therefore, whenever a bank uses transition matrices that were not estimated on internal portfolio data or where the data history could be insufficient, a basic validation of the stress test parameterization should be carried out. As a first step the portfolio $W_{ttc}$ should be computed and compared with a bank's current portfolio. Large deviations in some rating grades might already indicate potential instabilities in a credit risk stress test. In addition, a portfolio projection using the unstressed matrix ${\mathbf T}$ should be performed and average portfolio PD together with the evolution of the rating distribution should be inspected. This will allow a risk manager to detect whether spurious boom or recession periods are hidden in the stress test model parameters. If this is the case, the parameterization has to be improved before performing a credit risk stress test.

\section{Conclusion}

Credit risk stress tests are one of the most complex tasks in risk management. In a credit risk stress test, a bank's loan portfolio is projected over multiple years conditional on a macroeconomic scenario. While initially the stress test terms were limited to three years, more recently regulators require banks to perform stress over longer time horizons up to 30 years. This makes stress test even more demanding and increases the necessity of consistent parameters when executing a stress test. In practice, a key ingredient of a stress test is a rating transition matrix which in many cases cannot be estimated with high quality on internal portfolio data. In this case, transition matrices from rating agencies or matrices estimated on portfolios that are considered as representative are used as approximations.

It was demonstrated that this practice could lead to spurious stress testing results. It was shown that under mild assumptions a unique TTC portfolio exists that is independent of a bank's current portfolio and entirely determined by stress test model parameters. When a portfolio projection without stress is applied, the current bank portfolio will converge to the TTC portfolio. However, depending on the level of inconsistency between the current bank portfolio and the transition matrix, this convergence may not be monotonous. In these cases, the stress test could signal a recession or a boom which is not contained in the model but entirely an effect stemming from inconsistent model parameters.

For this reason, a risk manager should always perform a basic validation of stress testing model parameters before running a stress test. As outlined in Section \ref{sec:examples}, the TTC portfolio $W_{ttc}$ should be computed and compared with the current portfolio. Furthermore, projecting the current portfolio without stress and computing average PDs gives an indication whether a stress test might result in a spurious recession or underestimate the effects of a recession scenario. Although the examples in this article were computed over long time horizons, the impact of inconsistent parameters could be visible already in the first year of a stress test which means that every stress test could be impacted by this phenomenon. Therefore, to avoid or at least to anticipate these problems, the simple validations suggested in this article should be performed with every credit risk stress test.

\bibliography{sn-bibliography_2}

\appendix

\section{Proof of Theorem \ref{th:ttc}}

The proof of Theorem \ref{th:ttc} will be done by applying the Perron-Frobenius theorem \cite[Chapter 8]{meyer2023matrix}. To make the application of this theorem feasible, (\ref{eq:ttc}) has to be slightly transformed. Transposing (\ref{eq:ttc}) leads to
\begin{equation}
	W_{ttc} = {\mathbf I}_w {\mathbf T}' W_{ttc} + O V'_w {\mathbf T}' W_{ttc}.
\end{equation}
In the language of linear algebra, $W_{ttc}$ has to be an eigenvector of the matrix ${\mathbf M}$ given as
\begin{equation}
	{\mathbf M} = {\mathbf I}_w {\mathbf T}' + O V'_w {\mathbf T}'.
\end{equation}
with eigenvalue 1. It has to be shown that such an eigenvector exists, is unique and can be computed using the iterative algorithm of Theorem \ref{th:ttc}.

As a first step, it is shown that the dimension of the problem can be reduced by one. For this purpose, the matrix ${\mathbf M} = (m_{i,j})$ is computed in detail.
\begin{align*}
	{\mathbf M} & = \left(\begin{array}{ccccc} p_{1,1} + o_1p_{1,n} & p_{2,1} + o_1p_{2,n} & \cdots & p_{n-1,1} + o_1p_{n-1,n} & p_{n,1} + o_1 \\
		p_{1,2} + o_2p_{1,n} & p_{2,2} + o_2p_{2,n} & \cdots & p_{n-1,2} + o_2p_{n-1,n} & p_{n,2} + o_2\\
		\vdots   & \vdots   & \ddots & \vdots & \vdots \\
		p_{1,n-1} + o_{n-1}p_{1,n} & p_{2,n-1} + o_{n-1}p_{2,n} & \cdots & p_{n-1,n-1} + o_{n-1}p_{n-1,n} & p_{n,n-1} + o_{n-1} \\
		0 + o_np_{1,n} & 0 + o_np_{2,n} & \cdots & 0 + o_np_{n-1,n} & 0 + o_n \end{array}\right)
\end{align*}
In each expression for $m_{i,j}$, the part left of the plus sign comes from ${\mathbf I}_w {\mathbf T}'$ while the right part is the contribution of $O V'_w {\mathbf T}'$. By assumption, rating grade $n$ is absorbing which means $p_{n,1} = p_{n,2} = \cdots = p_{n,n-1} = 0$ and no loans are originated into grade $n$. Therefore, ${\mathbf M}$ is effectively a $n-1\times n$ matrix:
\begin{align*}
	{\mathbf M} & = \left(\begin{array}{ccccc} p_{1,1} + o_1p_{1,n} & p_{2,1} + o_1p_{2,n} & \cdots & p_{n-1,1} + o_1p_{n-1,n} & o_1 \\
		p_{1,2} + o_2p_{1,n} & p_{2,2} + o_2p_{2,n} & \cdots & p_{n-1,2} + o_2p_{n-1,n} & o_2\\
		\vdots   & \vdots   & \ddots & \vdots & \vdots \\
		p_{1,n-1} + o_{n-1}p_{1,n} & p_{2,n-1} + o_{n-1}p_{2,n} & \cdots & p_{n-1,n-1} + o_{n-1}p_{n-1,n} & o_{n-1} \\
		0 & 0 & \cdots & 0 & 0 \end{array}\right)
\end{align*}
By construction of the loan portfolio propagation, all defaulted balance is written off immediately and replaced by new origination into performing grades. Therefore, $W_{ttc}$ has to be of the form $W_{ttc} = (w_{ttc,1}, w_{ttc,2}, \ldots, w_{ttc,n-1}, 0)$. This reduces the problem of determining $W_{ttc}$ to a $n-1$-dimensional problem of finding an eigenvector $W_{ttc,p}$ with eigenvalue one of the reduced matrix ${\mathbf M}_p$ for the performing rating grades only:
\begin{align*}
	{\mathbf M}_p & = \left(\begin{array}{cccc} p_{1,1} + o_1p_{1,n} & p_{2,1} + o_1p_{2,n} & \cdots & p_{n-1,1} + o_1p_{n-1,n}\\
		p_{1,2} + o_2p_{1,n} & p_{2,2} + o_2p_{2,n} & \cdots & p_{n-1,2} + o_2p_{n-1,n}\\
		\vdots   & \vdots   & \ddots & \vdots \\
		p_{1,n-1} + o_{n-1}p_{1,n} & p_{2,n-1} + o_{n-1}p_{2,n} & \cdots & p_{n-1,n-1} + o_{n-1}p_{n-1,n}  \end{array}\right)
\end{align*}
By assumption ${\mathbf T}_p$ is a primitive matrix. Since ${\mathbf M}_p$ is constructed from ${\mathbf T}_p$ by transposing it and adding non-negative numbers, ${\mathbf M}_p$ has to be a primitive matrix, too. This means the Perron-Frobenius theorem can be applied to ${\mathbf M}_p$.

The Perron-Frobenius theorem implies that the matrix ${\mathbf M}_p$ has a positive real-valued eigenvalue $\lambda_{pf}$, the Perron-Frobenius eigenvalue. The multiplicity of this eigenvalue is one which means that the eigenvector is unique up to a scaling factor. Furthermore, all components of the eigenvector are strictly positive and sum to one after applying an appropriate scaling factor. For all other eigenvalues $\lambda$, the inequality $|\lambda| < \lambda_{pf}$ is fulfilled. For $\lambda_{pf}$ the inequality \[\min\limits_j\sum_{i=1}^{n-1}m_{p,i,j} \le \lambda_{pf}\le\max\limits_j\sum_{i=1}^{n-1}m_{p,i,j}\] holds. Since by construction all columns of ${\mathbf M}_p$ sum to one, $\lambda_{pf} = 1$.

Finally, to show that the iterative algorithm converges, decompose $W_{p,init}$ into a linear combination of the eigenvectors $v_1,v_2,\ldots,v_{n-1}$ where $v_1$ corresponds to the eigenvalue $\lambda_{pf} = 1$. For $v_1$ the components sum to one while for all other eigenvectors $v_2,\ldots,v_{n-1}$ the components sum to zero. Assume this for the moment as given, then every vector $W_{p,init}$ with components summing to one can be decomposed into $W_{p,init} = v_1 + \sum_{i=2}^{n-1}\beta_iv_i$. Then it holds
\begin{equation*}
	{\mathbf M}_p^k W_{p,init} = \lambda_1^kv_1 + \sum_{i=2}^{n-1}\beta_i\lambda_i^kv_i = v_1 + \sum_{i=2}^{n-1}\beta_i\lambda_i^kv_i \approx v_1,\ k\rightarrow\infty.
\end{equation*}
According to the Perron-Frobenius theorem, $\lambda_1 = 1$ and for all other eigenvalues $|\lambda_i| < 1$ which means that the sum is exponentially decaying to zero. To complete the proof, it will be shown that the components of all eigenvectors $v_2,\ldots,v_{n-1}$ sum to zero. Assume there is one eigenvector with components not summing to zero, e.g., the eigenvector $v_2$. After scaling, the components of $v_2$ sum to one. Now apply the iterative algorithm for $W_{p,init} = \gamma v_1 + (1-\gamma)v_2$ with $0 < \gamma < 1$. After one iteration
\begin{equation}
	{\mathbf M}_p W_{p,init} = \gamma v_1 + (1-\gamma)\lambda_2v_2 =: W_{p,1}.
\end{equation}
Since the Perron-Frobenius theorem states that $|\lambda_2| < 1$ the components of $W_{p,1}$ can no longer sum to one. This is a contradiction since by construction of the matrix ${\mathbf M}_p$ the sum of a vector's components is preserved under multiplication of this vector with ${\mathbf M}_p$.\ \qedsymbol

\end{document}